\documentclass[table]{ccjnl}
\usepackage{lipsum,amsmath}
\usepackage{cuted}
\usepackage{bm}
\usepackage{booktabs}
\usepackage{multirow}
\usepackage{makecell}
\usepackage{threeparttable}
\usepackage{booktabs}
\usepackage{array}
\graphicspath{{figures/}}
\title{AI Enlightens Wireless Communication: A Transformer Backbone for CSI Feedback }
\author{
Han Xiao\inst{1}, Zhiqin Wang\inst{\corinfo{zhiqin.wang@caict.ac.cn}*}, Dexin Li\inst{1}, Wenqiang Tian\inst{1}, Xiaofeng Liu\inst{2}, Wendong Liu\inst{1}, Shi Jin\inst{3}, Jia Shen\inst{1}, Zhi Zhang\inst{1}, Ning Yang\inst{1}
}
\receiveddate{}
\reviseddate{}
\Editor{}

\address[1]{Dept.of Standards Research, OPPO, Beijing, China}
\address[2]{China Academy of Information and Communications Technology, Beijing, China}
\address[3]{National Mobile Communications Research Laboratory, Southeast University, Nanjing, China}

\begin{document}

\maketitle

\begin{abstract}
This paper is based on the background of the 2nd Wireless Communication Artificial Intelligence (AI) Competition (WAIC) which is hosted by IMT-2020(5G) Promotion Group 5G+AI Work Group, where the framework of the eigenvector-based channel state information (CSI) feedback problem is firstly provided. Then a basic Transformer backbone for CSI feedback referred to EVCsiNet-T is proposed. Moreover, a series of potential enhancements for deep learning based (DL-based) CSI feedback including i) data augmentation, ii) loss function design, iii) training strategy, and iv) model ensemble are introduced. The experimental results involving the comparison between EVCsiNet-T and traditional codebook methods over different channels are further provided, which show the advanced performance and a promising prospect of Transformer on DL-based CSI feedback problem.
\end{abstract}

\section{Introduction}
\label{Introduction}
The physical layer is the basis for ensuring the quality of communication services for a wireless communication system. As a critical technology in physical layer for the fifth generation (5G), the massive multiple-input multiple-output (MIMO) will also be integral in the sixth generation (6G) to meet the growing needs of the mobile data traffic. Specifically, the channel state information (CSI) feedback is indispensable in massive MIMO to ensure that the base station (BS) can accurately obtain the channel quality and perform further resource allocation, data transmission, etc.

To fully explore the potential of massive MIMO, accurate CSI feedback can be conducted based on the codebook based mechanism including TypeI and enhanced TypeII (eTypeII), which have been standardized in release 16 (R16) by 3rd Generation Partnership Project (3GPP) \cite{1,2,3} and widely used for CSI feedback in current 5G new radio (NR) system. However, since the performance of such a scheme depends on the codebook, beam vector quantization error by codebook is inevitable, and the complexity of codebook design and the corresponding feedback overhead will increase significantly with the growing number of antennas and subcarriers, which brings considerable challenges to codebook based methods.

The core contributions are summarized as follows.
\begin{itemize}
\item With 2nd Wireless Communication Artificial Intelligence (AI) Competition (WAIC) as the background, a framework of eigenvector-based CSI feedback is introduced, where the system-level channel model for dataset and eigenvector-based system model provide a strong and effective guidance for future system design, performance evaluation and protocol specification in task of deep learning (DL) based CSI feedback.
\item Based on the provided framework, a Transformer backbone for CSI feedback referred to EVCsiNet-T is proposed, which processes all eigenvectors of the CSI as a squence using self-attention mechanism and can potentially be a plausible and convincing baseline for research of CSI feedback in academia and recent study item of 3GPP.
\item Beyond the backbone of EVCsiNet-T, a series of potential enhancements including data augmentation, loss function design, training strategy and model ensemble which can be exploited to further improve the performance of the backbone are introduced. 
\item To demonstrate the advanced performance and a promising prospect of Transformer on DL-based CSI feedback, we conduct various numerical experiments involving the comparison between the EVCsiNet-T and traditional codebook methods over different channels. 
\end{itemize}

The structure of this paper is as follows, the background including the relevant works and the 2nd WAIC with the task of DL-based eigenvector-based CSI feedback are introduced in Section \ref{Background}. The framework system model and channel model involved in the 2nd WAIC are depicted in Section \ref{System Model and Channel Model}. A basic Transformer backbone for CSI feedback named EVCsiNet-T is proposed in Section \ref{Basic Transformer Solution for CSI Feedback}. Moreover, a series of potential enhancements for DL-based CSI feedback is introduced in Section \ref{Enhancing Schemes}. Finally, experimental results and a brief conclusion are provided in Section \ref{Experiments} and \ref{CONCLUSION}, respectively.

\section{Background}
\label{Background}
\subsection{Relevant Works}
\label{Relevant Methods}
Since the DL has gained great successes in the fields of computer vision (CV) and natural language processing (NLP), the combination of wireless communication and DL has also attracted great attention in recent years \cite{Xiao2021AIEW, xiao2022channelgan, liu2021evcsinet}. The study item (SI) named `study on artificial intelligence (AI)/machine learning (ML) for NR air interface` has been established in 3GPP release 18 (R18) in which DL-based CSI feedback is regarded as an important use case \cite{213599, 212927, 210235, 210236}. In academia, DL-based CSI feedback has been widely studied since it can simultaneously achieve higher compression and recovery accuracy and lower feedback overhead \cite{wang2017deep,wen2018deep,sun2020ancinet,lu2020multi,chen2020deep,mashhadi2020distributed,cao2021lightweight,guo2020deep,guo2021canet,lu2018mimo,wang2018deep,guo2020convolutional,li2020spatio,chen2019novel,lu2019bit}. 
CsiNet \cite{wen2018deep} is firstly proposed whose encoder and decoder are constructed by convolutional neural network (CNN) to conduct CSI compression and recovery at the user equipment (UE) and BS, respectively. Furthermore, a series of follow-up studies designed based on various kinds of CNNs \cite{sun2020ancinet,lu2020multi,chen2020deep,mashhadi2020distributed,cao2021lightweight,guo2020deep,guo2021canet} and  long short-term memory (LSTM) \cite{lu2018mimo} have been proposed to further improve the CSI feedback performance. It can be noticed that most studies originated from CsiNet mainly focus on the full channel state information (F-CSI) feedback, where the whole downlink channel matrix is compressed and recovered at encoder and decoder, respectively. However, the solution for CSI feedback discussed in 3GPP is based on the compression and feedback of the eigenvector of the channel matrix, which is the main application in current 5G system. Therefore, besides the DL-based F-CSI feedback, it is necessary to study the eigenvector-based CSI feedback using DL methods, which can achieve direct fair performance comparison with the existing codebook based solutions. Different from the DL-based F-CSI feedback, EVCsiNet \cite{liu2021evcsinet} is introduced which concentrates on eigenvector-based CSI feedback. However, the current research on eigenvector-based CSI feedback is still relatively preliminary calling for more exploration, which thus is the main focus of this paper.

\subsection{Wireless Communication AI Competition}
\label{WAIC2nd}
In order to further explore the practical application of AI in wireless communication systems, IMT-2020(5G) Promotion Group 5G+AI Work Group held the 2nd WAIC in July 2021 with a topic of AI Enlightens Wireless Communication which is still committed to promoting the deep integration and mutual promotion of the wireless communication and AI. The 2nd WAIC focuses on the task of Performance Improvement of DL-based CSI feedback and DL-based channel estimation. As one of the tasks, DL-based CSI feedback aims to evaluate the performance of DL-based CSI recovery and feedback overhead reduction with classic data-set on 3GPP system-level channel model. Since the high degree of integration between the task and the industry, there are more than $300$ teams involving the contestants from related companies, universities and research institutes participate in the competition. Focusing on the schedule of the competition, the two-month online results submission for participants and the one-day offline seminar for interested researchers provide a good opportunity for technical exploring and sharing.

As for the task of DL-based CSI feedback in 2nd WAIC, the participants achieve excellent results and there are six over top ten teams utilize the Transformer \cite{Vaswani2017AttentionIA} architecture as the backbone for model design. The Transformer architecture was firstly proposed to solve problems in the field of natural language processing (NLP) and computer vision (CV) \cite{Dosovitskiy2021AnII}, and the successful utilization in 2nd WAIC shows great potential of Transformer in the DL-based CSI feedback task.

In the following, based on the background of 2nd WAIC, the details of the i) dataset construction method, ii) the Transformer solution for CSI feedback and iii) the noteworthy enhancing schemes for CSI feedback are introduced. The corresponding experimental results are also provided. With the success of the competition, it becomes an unprecedented way to promote the integration of 5G and AI by involving academic and industrial research and development to solve typical wireless problems with AI tools.

\begin{table}[tbp]\small
\centering
\caption{Dataset parameter settings for 2nd WAIC.}
\label{tabChannelSeting}
\begin{tabular}{cc}
\hline
 Parameter  & Value \\
\hline
Channel model  &  UMa \& NLoS\\
 \hline
Carrier frequency $F_{\rm c}$ &  3.5GHz\\
 \hline
Subcarrier spacing  $B_{\rm sc}$  &  15KHz\\ 
 \hline
Number of resource blocks $N_{\rm RB}$  &  48\\
\hline
Number of subbands $N_{\rm sb}$  &  12\\
\hline
 Number of Tx antennas $N_{\rm t}$ & 32 \\
 \hline
 Number of Rx antennas $N_{\rm r}$ & 4\\
 \hline
 UE speed  & 3km/h\\
 \hline
 Number of cells $N_{\rm c}$ & 57\\
 \hline
 Number of UEs in training set $N_{\rm train}$  &  3000 \\
 \hline
 Number of UEs in testing set $N_{\rm test}$  & 400 \\
 \hline
 Number of sampling slots $N_{\rm slot}$  &  200 \\
 \hline
 Number of interval slots $T$  &  50\\
 \hline
\end{tabular}
\end{table}

\section{Framework of Eigenvector-Based CSI Feedback}
\label{System Model and Channel Model}
\subsection{Channel Model}
\label{Channel Model}
3GPP technical report (TR) 38.901 \cite{4} entitled `Study on channel model for frequencies from 0.5 to 100 GHz` helps to properly model and evaluate the performance of physical layer techniques using the appropriate channel model. Since the channel models in TR 38.901 are recognized and typical, the study of CSI feedback based on these channel models is appropriate and persuasive. As a benchmark channel for simulation and performance evaluation in 3GPP, the system level three-dimensional channel model is adopted in this paper to reflect the channel characteristics in realistic environments. Specifically, four types of channel models including urban-macro (UMa), urban-micro indoor-hotspot, and rural-macro are defined in TR 38.901, and we selectively consider the UMa model with pure non-line-of-sight (NLoS). The system-level time domain downlink channel matrix $\mathbf{H}_{\rm{t}}$ can be written as
\begin{equation}
\mathbf{H}_{\rm{t}} = \sum_{d=1}^{N_{\rm d}} \widetilde{\mathbf{H}}_d = \sum_{d=1}^{N_{\rm d}}\sum_{l=1}^{L_d} \widetilde{\mathbf{H}}_{d,l}
\label{eqHsys}
\text{,}
\end{equation}
where $\widetilde{\mathbf{H}}_{d,l}$ and $L_d$ denote the channel of $l$-th sub-path and the total number of sub-paths in the $d$-th cluster, respectively. Note that the channel defined in TR 38.901 is considered, where each sub-path $\widetilde{\mathbf{H}}_{d,l}$ is generated with different angle-of-departures, angle-of-arrivals, angle-spreads in both azimuth and zenith domains, power and delay distributions, and initial phases, etc, therefore the small-scale fading of frequency selective also occurs. Moreover, due to the frequency selection characteristics of the channel, CSI of different subcarriers are different, which introduces the subband-based CSI feedback. Moreover, $N_{\rm slot}\times N_{\rm UE}$ channel samples are provided in related dataset, where $N_{\rm slot}$ slots are uniformly sampled with $T$ continuous interval slots for each UE, $N_{\rm UE} = N_{\rm train} + N_{\rm test}$ denotes the number of randomly distributed UEs in $N_{\rm c}$ cells, $N_{\rm train}$ and $N_{\rm test}$ represent the numbers of UEs in training  and testing set, respectively. The corresponding channel and dataset parameter settings in 2nd WAIC are shown in Table \ref{tabChannelSeting}. Note that in order to balance the difficulty of the competition for the contestants and make the score distribution of the leaderboard reasonable, frequency selective fading of the channel is further reduced. The dataset (Data-WAIC2nd.zip) and an example code (reference model\_WAIC2nd\_EVCsiNet-T.zip) involved in 2nd WAIC are released in the page of 'News' of Wireless-Intelligence \footnotemark[1]\footnotetext[1]{https://wireless-intelligence.com/} which is a channel database website provided for AI-based wireless communication research. Since the dataset are constructed and  modified based on 3GPP specification, the results on this channel dataset have strong guiding significance for further research on the future communication system design and protocol specification.

\begin{figure}[tb]
\centering
\includegraphics[scale=0.55]{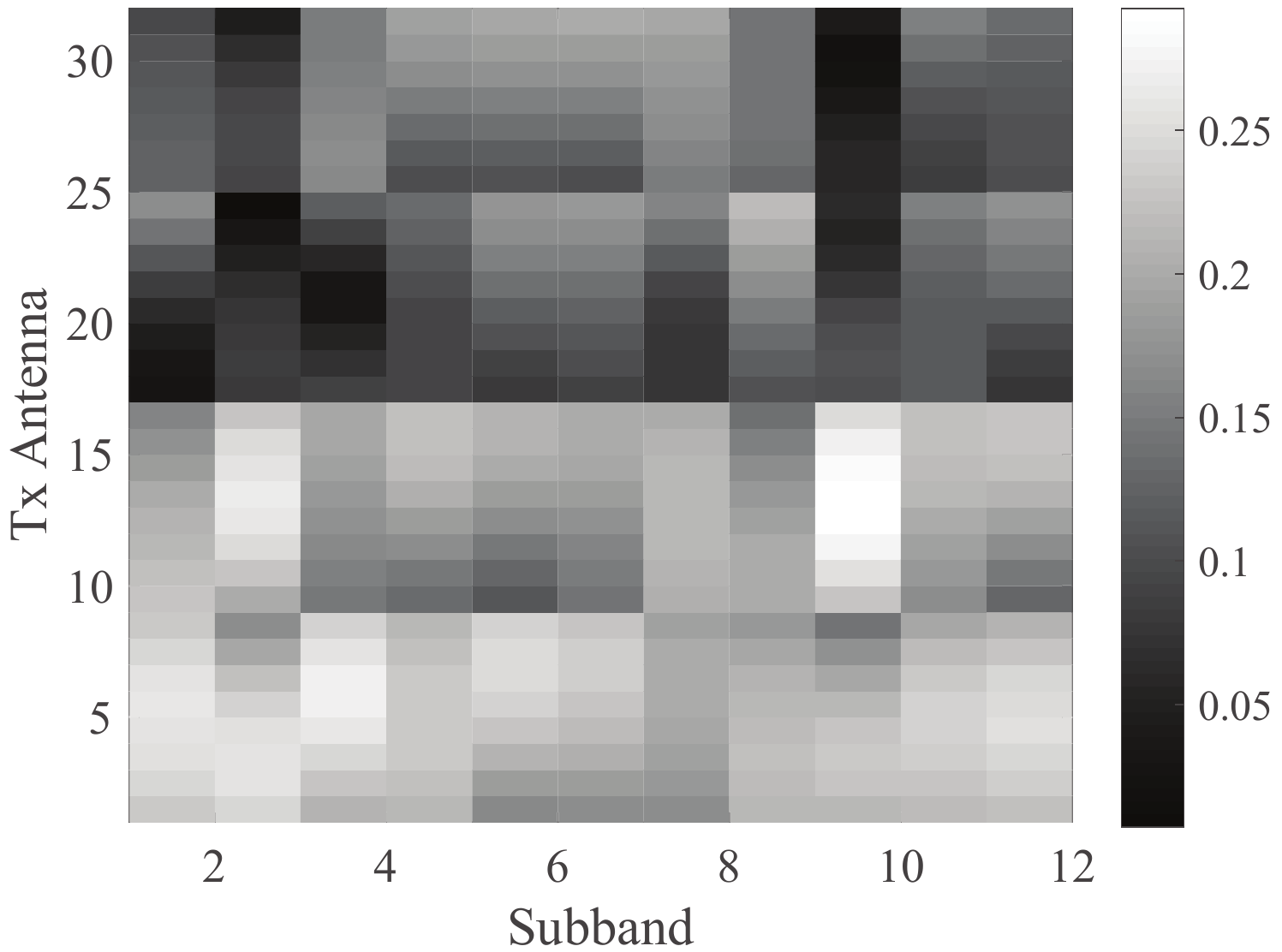}
\caption{Illustration of gray scale of energy for one sample.}
\label{gray}
\end{figure}

\subsection{System Model}
\begin{figure}[tb]
\centering
\includegraphics[scale=0.8]{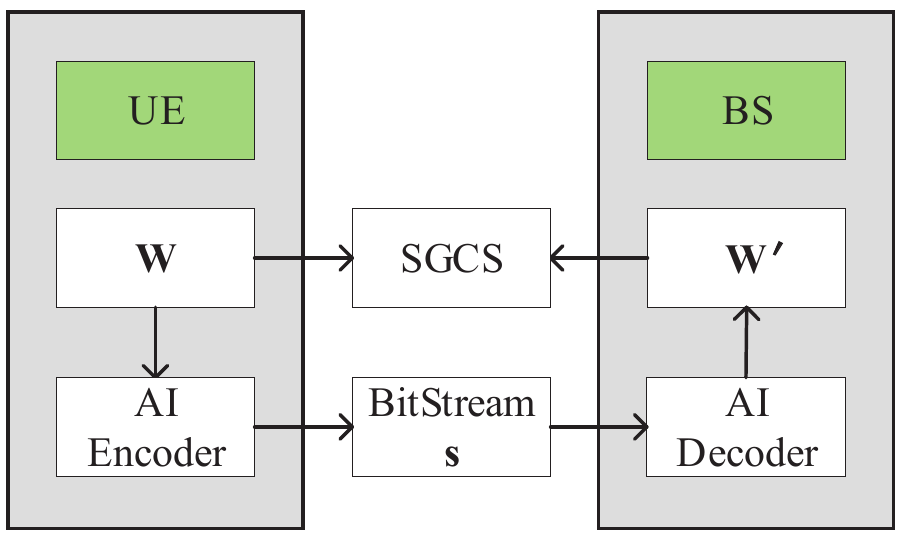}
\caption{Illustration of DL-based E-CSI feedback scheme.}
\label{basicScheme}
\end{figure}

We consider a typical MIMO system with $N_{\rm t}$ transmitting antennas at BS and $N_{\rm r}$ receiving antennas at UE. Due to the frequency selection characteristics of the channel, CSI of different subcarriers are different, which introduces the subband-based CSI feedback by dividing the full bandwidth into multiple subbands for feedback. By conducting fast Fourier transform (FFT) on the time domain downlink channel matrix  $\mathbf{H}_{\rm{t}}$ described in Section \ref{Channel Model}, the downlink channel in the frequency domain can be given as
\begin{equation}\label{Hf}
\mathbf{H}_{\rm{f}}= \big[\mathbf{H}_1,\mathbf{H}_2,\cdots,\mathbf{H}_{N_{\rm{sb}}}\big],
\end{equation}
where $N_{\rm{sb}}$ represents the number of subbands consisting of 4 resource blocks (RBs) as the basic feedback granularity, and $\mathbf{H}_k\in\mathbb{C}^{N_{\rm r}\times N_{\rm t}}, 1\leq k\leq N_{\rm{sb}}$ indicates the downlink channel of the $k$th subband. According to the application of the current 5G system, the solution for CSI feedback is based on the compression and feedback for the eigenvector of the channel matrix. However, most existing academia researches \cite{wang2017deep,wen2018deep,sun2020ancinet,lu2020multi,chen2020deep,mashhadi2020distributed,cao2021lightweight,guo2020deep,guo2021canet,lu2018mimo,wang2018deep,guo2020convolutional,li2020spatio,chen2019novel,lu2019bit}  for DL-based CSI feedback mainly concentrate on full CSI compression and recovery, which is incomparable with the current Type I and eType II codebook based mechanisms with eigenvector CSI operation defined by 3GPP. Moreover, since the study item named 'study on artificial intelligence (AI)/machine learning (ML) for NR air interface' has been established in 3GPP R18 in which DL-based CSI feedback is regarded as an important use case \cite{213599, 212927, 210235, 210236}, the study for advanced structure of neural network (NN) and comparison between DL-based methods and codebook based methods are significant and indispensable. These motivate the study of this paper, and the DL-based CSI feedback for eigenvector considered in this paper can also achieve fair comparison with the traditional codebook based mechanism including TypeI and eTypeII in the 5G system. Assuming ideal channel estimation at UE and implementing the single layer downlink transmission, the corresponding eigenvector for the $k$th subband $\mathbf{w}_k\in\mathbb{C}^{N_{\rm t}\times 1}$ with normalization $||\mathbf{w}_k||^2=1$, can be utilized as the downlink precoding vector and calculated using eigenvector decomposition, i.e.,
\begin{equation}
\mathbf{H}_k^{\rm H}\mathbf{H}_k \mathbf{w}_k =\lambda_k \mathbf{w}_k,
\end{equation}
where $\lambda_k$ represents the maximum eigenvalue of $\mathbf{H}_k^{\rm H}\mathbf{H}_k$. As all $N_{\rm{sb}}$ eigenvectors should be feeded back to the BS for downlink precoding, total $N_{\rm{sb}}N_{\rm t}$ complex coefficients for $N_{\rm{sb}}$ subbands should be compressed and recovered using NN. Thus a CSI sample $i$ across total $N_{\rm{sb}}$ subbands can be writen as
\begin{equation}\label{Wi}
\mathbf{W}_i= \big[\mathbf{w}_{1,i},\mathbf{w}_{2,i},\cdots,\mathbf{w}_{N_{\rm{sb}},i}\big] \in \mathbb{C}^{N_{\rm t}\times N_{\rm{sb}}}.
\end{equation}
The gray scale of $\mathbf{W}$ is also described in Fig. \ref{gray}, where x-axis indicates the subband index and y-axis indicates the corresponding Tx antenna index, and deeper color means the smaller energy of the corresponding channel element. For CSI compressing and feedback, a DL-based CSI feedback scheme is implemented as shown in Fig. \ref{basicScheme}, where a CSI sample $\mathbf{W}$ is compressed to a bitstream $\mathbf{s}$ of length $M$ using a DL encoder at UE. Then the CSI $\mathbf{W}$ is recovered from the feedback bitstream $\mathbf{s}$ exploiting the DL decoder at the BS. Moreover, denoting the test set $\mathcal{W}$ consisting of $|\mathcal{W}|$ CSI samples where $|\cdot|$ denotes the cardinality of a set, the average squared generalized cosine similarity (SGCS) is usually utilized to evaluate the CSI compression and recovery accuracy as follows

\begin{equation}\label{score_function}
\setlength{\abovedisplayskip}{-5pt}
\setlength{\belowdisplayskip}{5pt}
\begin{split}
\rho(\mathcal{W},\mathcal{W}') = \frac{1}{|\mathcal{W}|N_\textrm{sb}}\sum_{i=1}^{|\mathcal{W}|}\sum_{k=1}^{N_\textrm{sb}}\Big(\frac{\|\mathbf{w}_{k,i}^{\rm H}\mathbf{w}'_{k,i}\|_2}{\|\mathbf{w}_{k,i}\|_2\|\mathbf{w}'_{k,i}\|_2}\Big)^2
\end{split}
\text{,}
\end{equation}
where $\rho(\mathcal{W},\mathcal{W}')\in [0,1]$, $\|\cdot\|_2$ denotes the $\ell_{2}$ norm, $\mathcal{W}'$ represents the predicted eigenvector CSI set, $\mathbf{w}_{k,i}$ and $\mathbf{w}'_{k,i}$ represent the $k$th eigenvectors of $i$th sample in $\mathcal{W}$ and $\mathcal{W}'$, respectively. A higher SGCS performance indicates higher CSI compression and recovery accuracy. Based on the basic requirements of wireless communication, the 2nd WAIC focuses on the SGCS performance under different feedback overhead in a complex channel environment. The model design with $M = 48$ and $M = 128$ bits are required which correspond to low feedback overhead scenario and high feedback overhead scenario, respectively. The final score is obtained by averaging the SGCS performance of the two scenarios.

\section{A Transformer Backbone for CSI Feedback}
\label{Basic Transformer Solution for CSI Feedback}
Recently, Transformer \cite{Vaswani2017AttentionIA} based solution shows great potential in CSI feedback, whose natural encoder-decoder structure is quite suitable for respectively implementing at the UE and BS. By implementing Transformer for CSI feedback, a specific domain of the channel such as paths, antenna pairs, or subbands can be treated as the sequence inputs and processed by self-attention block, which is similar to the sequential processing issue in NLP and benefits the extraction of unique characteristics of the wireless channel. Here, a Transformer backbone referred to EVCsiNet-T for solving CSI feedback is introduced, where all subbands are regarded as a sequence. By splitting the real and imaginary parts of a complex CSI sample, the input of the EVCsiNet-T can be written as 
\begin{equation}
\widehat{\mathbf{W}} = [{\rm Re}(\mathbf{W})^{\rm T},{\rm Im}(\mathbf{W})^{\rm T}]^{\rm T} \in \mathbb{R}^{2N_{\rm t}\times N_{\rm{sb}}},
\end{equation}
where the subscript $i$ indicating the index of a sample is omitted to simplify the representation, and ${\rm Re}(\cdot)$ and ${\rm Im}(\cdot)$ denote the real and imaginary parts of a matrix, respectively. To implement the CSI feedback, the EVCsiNet-T solves the following problem during the training phase, i.e.,
 \begin{equation}
\begin{split}
\mathop{\min}_{\Theta_{\rm E},\Theta_{\rm D}} L(\widehat{\mathbf{W}},\widehat{\mathbf{W}}')
\end{split}
\text{,}
\end{equation}
where $\Theta_{\rm E}$ and $\Theta_{\rm D}$ denote the weights of encoder and decoder of EVCsiNet-T, respectively, and $L(\cdot)$ denotes the loss function such as mean square error (MSE), minus cosine similiarity, etc.

\begin{figure}[tb]
\centering
\includegraphics[scale=0.8]{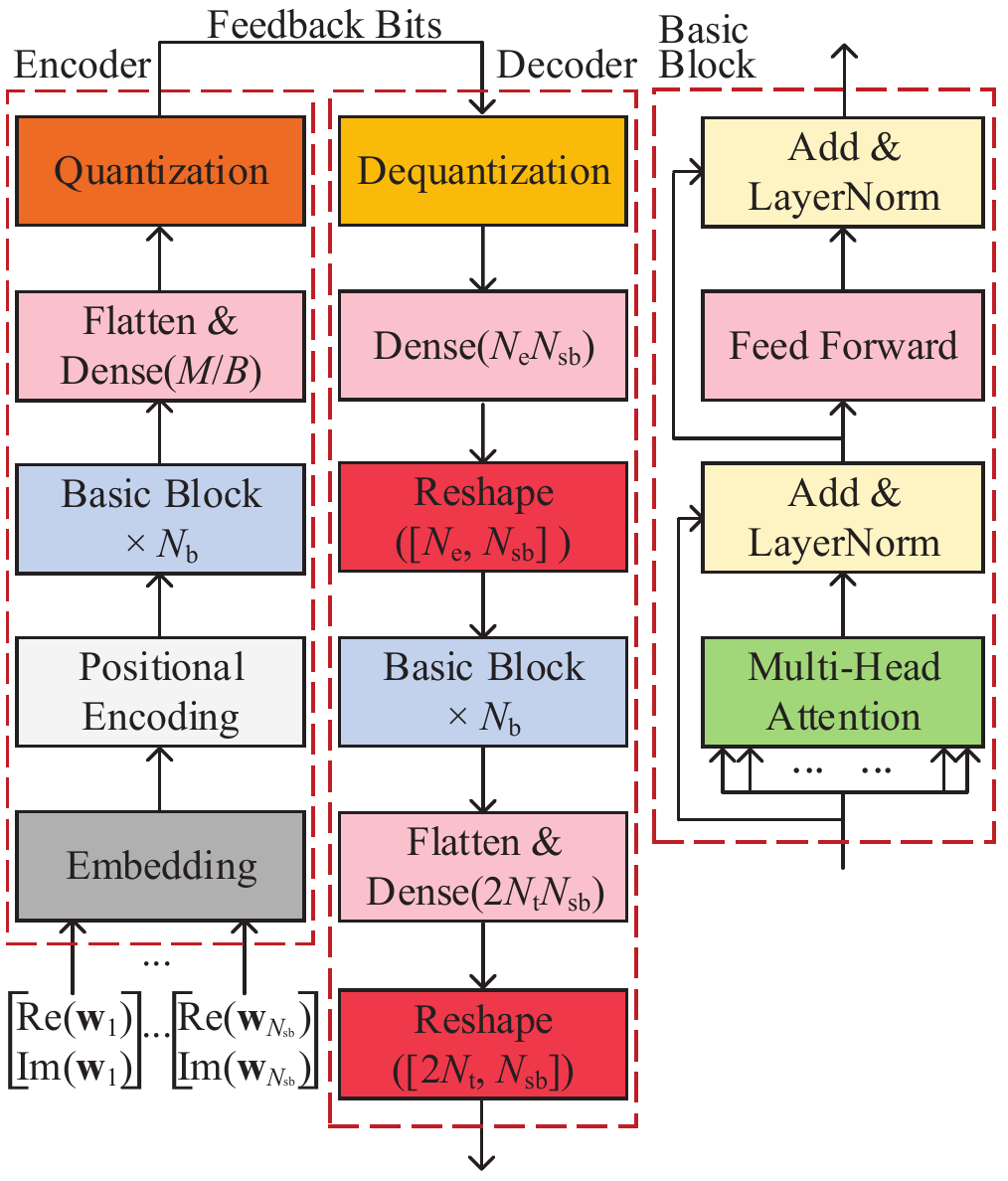}
\caption{Illustration of architecture of EVCsiNet-T.}
\label{basicTransformer}
\end{figure}

The architecture of EVCsiNet-T is depicted in Fig. \ref{basicTransformer} in detail which consists of encoder and decoder. Each vector of subband in the input CSI matrix is firstly processed by an embedding layer with embedding dimension of $N_{\rm{e}}$ and positionally encoded by adding a learnable position vector. Then $N_{\rm{b}}$ basic blocks are sequentially introduced to conduct the feature extraction. Moreover, a flatten layer and a dense layer with $M / B$ units are implemented in order to adapt the length of feedback bits, where $B$ is the number of quantization bits and $M$ denotes the total length of feedback bits. Finally, a uniform quantization layer is employed to transform the floating number vector with length of $M / B$ to the bitstream.

As for the decoder, the fed back bitstream is converted to a vector of float numbers using the dequantization layer. Next, a dense layer with $N_{\rm{e}}N_{\rm{sb}}$ units using gaussian error linear units (GELU) activation function is employed, whose output is reshaped to a $N_{\rm{e}} \times N_{\rm{sb}}$ matrix. After that, $N_{\rm{b}}$ basic blocks are sequentially deployed to extract the features, whose output is flattened and processed by a dense layer with $2N_{\rm{t}}N_{\rm{sb}}$ units and activation function of GELU. Finally, the reshape layer is implemented to obtain the output with the shape of the original CSI.

The basic block can be the core component in both encoder and decoder, in which the input is firstly processed by a multi-head attention layer \cite{Vaswani2017AttentionIA} with the number of heads $N_{\rm{head}}$ and then added to the shortcut from the input of the block. A layer normalization and feedforward network are sequentially conducted where the feedforward network is constructed by two dense layers with the number of units $\{ k_{\rm h}N_{\rm{e}}, N_{\rm{e}}\}$ sandwiching a GELU, where $k_{\rm h}$ denotes the scaling factor of hidden layer. The shortcut from the input of feed forward network is also added to the output and a layer normalization is implemented finally.

\section{Potential Enhancements}\label{Enhancing Schemes}
Different from the image data and sequence data respectively in CV and NLP, the channel data obtained from wireless communication system has some unique features, which call for more suitable tactics for performance enhancing. Specifically, based on the backbone introduced in Section \ref{Basic Transformer Solution for CSI Feedback}, there are also a series of enhancing schemes for DL-based CSI feedback during the 2nd WAIC, which can further improve the performance of the model and are introduced in more detail in this section.

\subsection{Data Augmentation}\label{Data Augmentation}

The data augmentation approach is widely utilized to improve the diversity of the training data and alleviate the overfitting problem. There are a number of data augmentation approaches exploited in the competition including noise injection, flipping, shift, and rotation, which are introduced in detail in this subsection.

\subsubsection{Noise Injection}

Noise injection refers to adding noise to the training data which is shown in Fig. \ref{flipAndShift} (a). The noise is typically drawn from the various statistical distributions. In general, the noise is injected with a zero-mean Gaussian distribution, under which condition the injection process can be mathematically written as

\begin{equation}\label{noise_injectioin}
\mathbf{W}_{\rm{aug}} =
\mathbf{W} + \alpha \Upsilon, \quad \eta \sim \mathcal{N}(0,\sigma^2),
\end{equation}
where $\mathbf{W}_{\rm{aug}}$ denotes the augmented CSI tensor, $\mathbf{W}$ denotes the noise-free CSI tensor, $\Upsilon$ represents the noise tensor with the same shape as the CSI tensor $\mathbf{W}$ and with elements sampled from a Gaussian distribution $\mathcal{N}(0,\sigma^2)$, and $\alpha$ is the coefficient that scales the magnitude of the injected noise, respectively.

\subsubsection{Flipping}
\begin{figure}[tb]
\centering
\includegraphics[scale=0.6]{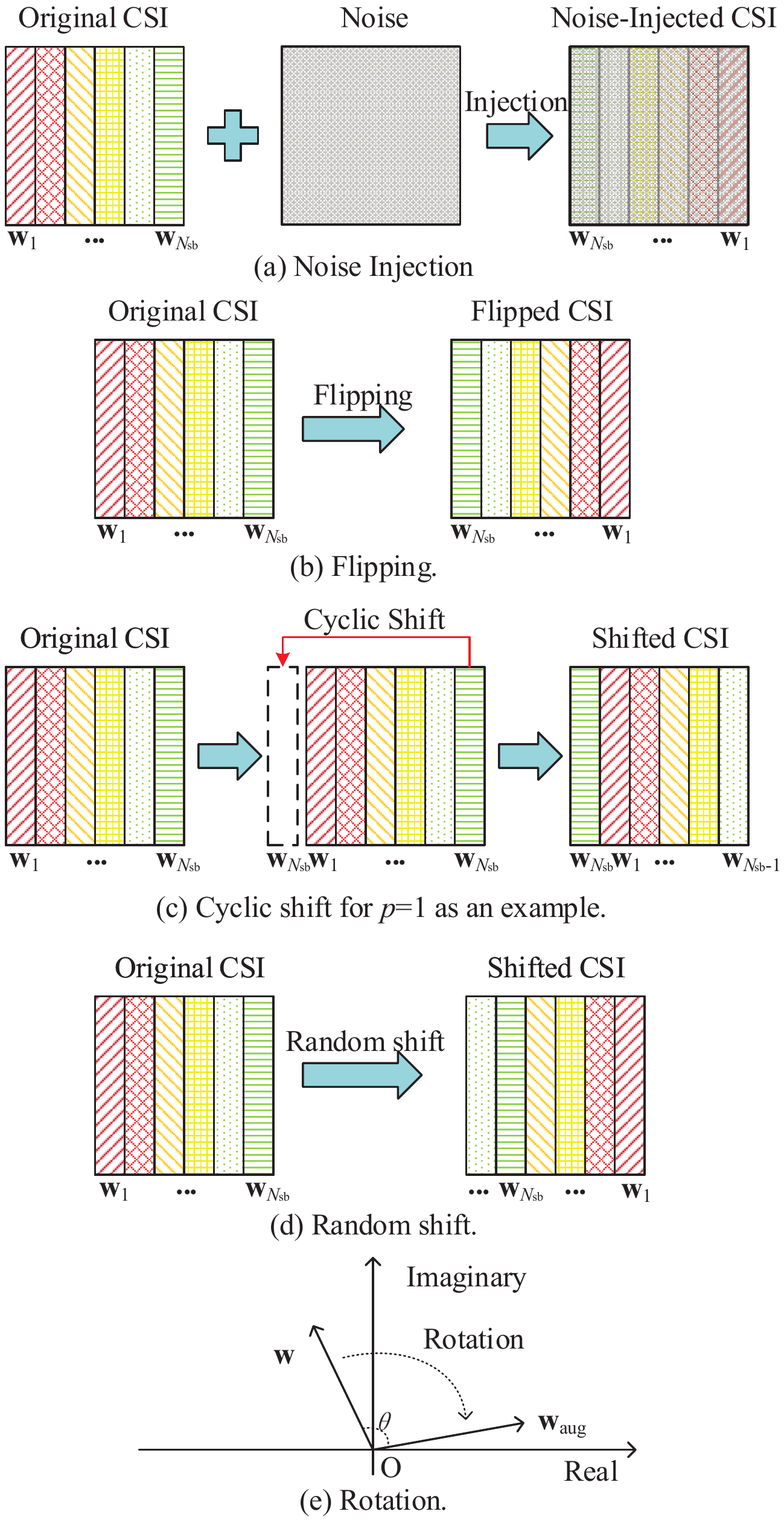}
\caption{Data augmentation schemes.}
\label{flipAndShift}
\end{figure}

The flipping method for CSI considers flip the CSI tensor according to the dimension of subband or antenna pairs. As shown in Fig. \ref{flipAndShift} (b), for a CSI tensor
\begin{equation}
\mathbf{W}= \big[\mathbf{w}_{1},\mathbf{w}_{2},\cdots,\mathbf{w}_{N_{\rm{sb}}}\big] \in \mathbb{C}^{N_{\rm t}\times N_{\rm{sb}}}
\end{equation}
as an example, where $\mathbf{w}_k\in\mathbb{C}^{N_{\rm t}\times 1}$ denotes the corresponding eigenvector for the $k$th subband, the flipping procesure according to the dimension of subband can be written as
\begin{equation}\label{aug_Flipping}
\mathbf{W}^{\rm{aug}} = \big[\mathbf{w}_{N_{\rm{sb}}},\mathbf{w}_{N_{\rm{sb}-1}},\cdots,\mathbf{w}_{1}\big] \in \mathbb{C}^{N_{\rm t}\times N_{\rm{sb}}},
\end{equation}
where $\mathbf{W}_{\rm{aug}}$ denotes the augmented CSI tensor and flipping in antenna paire dimensions can be easily generalized..

\subsubsection{Shifting}
Shifting method considers to shift along a certain dimension of the tensor according to the step size, which can change the position of the content of the tensor. Fig. \ref{flipAndShift} (c) shows the cyclic shifting of a CSI tensor $\mathbf{W}$ on the dimension of subband. It shifts the last subband to the first position for $p$ times, which can be written as
\begin{equation}\label{aug_circular_shift}
\mathbf{W}^{\mathbf{aug}}= \big[\mathbf{w}_{N_{\rm{sb}}-p+1}, \cdots,\mathbf{w}_{N_{\rm{sb}}}, \mathbf{w}_{1},\cdots,\mathbf{w}_{N_{\rm{sb}}-p}\big],
\end{equation}
where $\mathbf{w}_k\in\mathbb{C}^{N_{\rm t}\times 1}$ denotes the corresponding eigenvector for the $k$th subband and $N_{\rm sb}$ denotes the number of subbands. Different from the cyclic shifting, there can also be another method refered to random shift achieves data augmentation by randomly shuffling the subbands, which is shown in Fig. \ref{flipAndShift} (d). In more detail, cyclic shift can generate at most $N_{\rm sb} - 1$ samples based on one raw data sample while random shift does at most $N_{\rm sb}! - 1$ samples. Therefore random shift can provide more diversity of data augmentation than cyclic shift. However, since the relationship of adjacent subbands does not change when cyclic shifting, thus certain frequency selectta ive fading features are still retained in the generated data. 

\subsubsection{Rotation}
 
Rotation augmentation shown in Fig. \ref{flipAndShift} (e) utilizes sine and cosine functions to randomly rotate the eigenvector by a certain angle, which can be written as
\begin{equation}
\begin{split}
\operatorname{Re}\left\{\mathbf{w}_{\rm aug}\right\} =\cos\left(\theta\right)\operatorname{Re}\left\{\mathbf{w}\right\} -\sin\left(\theta\right)\operatorname{Im}\left\{\mathbf{w}\right\} ,
\\
\operatorname{Im}\left\{\mathbf{w}_{\rm aug}\right\}
=\sin\left(\theta\right)\operatorname{Re}\left\{\mathbf{w}\right\}  +\cos\left(\theta\right)\operatorname{Im}\left\{\mathbf{w}\right\} ,
\end{split}
\end{equation}
where $\operatorname{Re}\{\cdot\}$ and $\operatorname{Im}\{\cdot\}$ are the real and imaginary part of the input, $\mathbf{w}$ is one subband eigenvector of the CSI tensor $\mathbf{W}$, $\theta$ is the rotation degree parameter which can be different for defferent eigenvectors, respectively.
%Combining augmentations such as flipping, shift, and rotation can result in massively inflated dataset sizes. However, this is not guaranteed to be advantageous. 

\subsection{Loss Function Design}
\label{Loss Function Design}
Loss function design is a significant factor affecting the training. The minus cosine similarity and MSE which are commonly used always fail to strictly satisfy the final goal or unique features of the task, calling for novel method for designing the loss function. Here, we will give an introduction on enhanced loss function design for DL-based CSI feedback.

\subsubsection{Scoring Loss}
Since the final goal of the task is the pursuit of a high score according to (\ref{score_function}), a design method can be directly adapt the scoring indicator function as the loss function, i,e.,
\begin{equation}
L = -\rho,
\end{equation}
where $\rho$ is the scoring function in (\ref{score_function}) to evaluate the accuracy of the CSI compression and recovery.

\subsubsection{Quantization Error Compensation Loss}

The quantization error can be up to half the quantization interval and are always unavoidable. Therefore the quantization error compensation loss $L_{\rm quant}$ can be designed to reduce the error between the vector before quantization $\mathbf{v}$ and the vector after dequantization $\mathbf{v}'$ caused by quantization procedure, i.e.,
\begin{equation}
L_{\rm{quant}} = L_{\rm{base}} (\mathbf{v}, \mathbf{v}'),
\end{equation}
where $L_{\rm{base}}$ is differentiable and can be implemented using MSE, normalized MSE (NMSE) and so on, and $\mathbf{v}$ and $\mathbf{v}'$ are the input and output vectors of the quantization layer, respectively. Note that the quantization layer consisting of quantization and dequantization processes is generally non-differentiable in mathematics, but can still be implemented during training where same gradient is applied before and after the quantization layer. The quantization error compensation loss can be added to the original loss function forming the global loss to enhance the recovery performace, i.e.,
\begin{equation}
L = L_{\rm{original}} +L_{\rm{quant}},
\end{equation}
where $L_{\rm{original}} = L_{\rm{base}}(\widehat{\mathbf{W}},\widehat{\mathbf{W}}')$ reduces the error between the original CSI $\widehat{\mathbf{W}}$ and the recovered CSI $\widehat{\mathbf{W}}'$, $L_{\rm{quant}}$ reduces the quantization error to further improve the CSI recovery accuracy, and $L_{\rm{base}}$ can also be defined as MSE, NMSE, etc.

\subsection{Training Strategy}
\label{Tricks for Training}
\subsubsection{Learning Rate Warm-up and Decay}
Since the weights of the model are randomly initialized, a large learning rate may bring instability and oscillation of the model at the beginning of training. The warmup method can make the learning rate small during the first several training epochs, where the model can gradually become stable. After that, the training can continue under a learning rate set according to certain strategies leading to a better convergence of the model. One of the prevalent learning rate warmup methods is the gradual warmup method, using which the learning rate is updated by
\begin{equation}
\alpha_t = \dfrac{t}{T_{\rm warmup}}\alpha_{\rm max},
\end{equation}
where $t \in [1,T_{\rm warmup}]$ denotes the $t$th epoch, $T_{\rm warmup}$ denotes the number of warm-up epochs, and $\alpha_{\rm max}$ denotes the final warmed up learning rate, respectively.

After warm-up, learning rate decay can be further utilized which starts with the warmed up learning rate $\alpha_{\rm max}$ and then decaying it $T_{\rm decay}$ epochs in subsequent training. This can be beneficial for the optimization and generalization of the model \cite{You2019HowDL}. As an example, the cosine decay strategy can be written as
\begin{equation}
\begin{split}
\alpha_{t}=\alpha_{\min }&+\frac{1}{2}\left(\alpha_{\max }-\alpha_{\min }\right)\\
&\cdot\left(1+\cos \left(\frac{(t-T_{\rm warmup}) \pi}{T_{\rm decay}}\right)\right),
\end{split}
\end{equation}
where $t \in [T_{\rm warmup},T_{\rm decay}]$ denotes the $t$th epoch and $\alpha_{\min}$ represents the minimum values of the learning rate, respectively.

\begin{figure}[tb]
\centering
\includegraphics[scale=0.85]{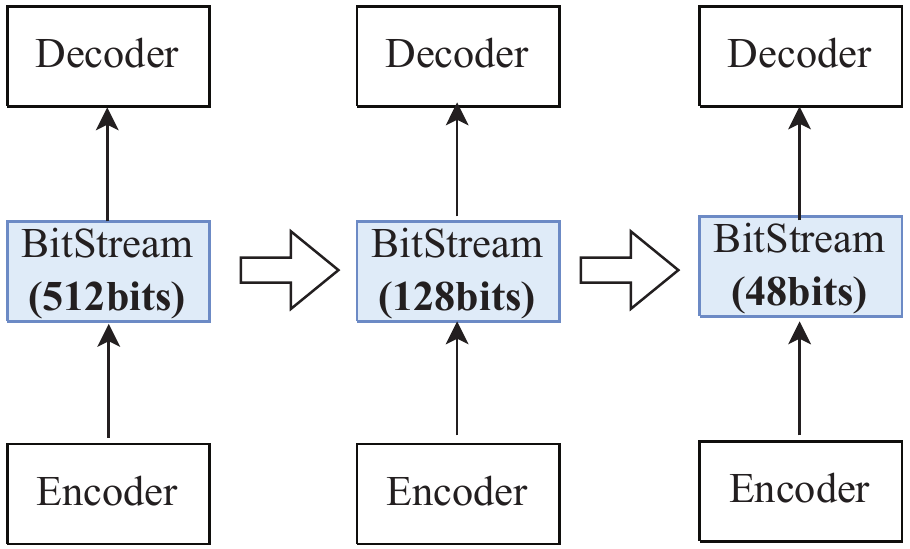}
\caption{Staged training for shrinking the number of feedback bits in different stages.}
\label{Progressive_Resizing}
\end{figure}

\subsubsection{Staged Training}
Another training strategy referred to staged training considers dividing the entire training procedure into several stages. where different factors, e.g., loss function, batch size, learning rate, datasets, can be adjusted in different training stages. Hence the design of the staged training is highly personalized. For example, progressive resizing is a powerful staged training method in CV for improving the training efficiency and effect, which gradually expands the size of image in consequent training stages \cite{Liang2021GuidanceNW}. Similar idea can also be applied in DL-based CSI feedback. As shown in Fig. \ref{Progressive_Resizing}, one can train with a large number of feedback bits for certain epochs and then shrinks to smaller feedback bits in the next re-training stage, in which way a more effective and stable convergence can be obtained.

\subsection{Model Ensemble}
\label{Model Ensembling}
\begin{figure}[tb]
\centering
\includegraphics[scale=0.78]{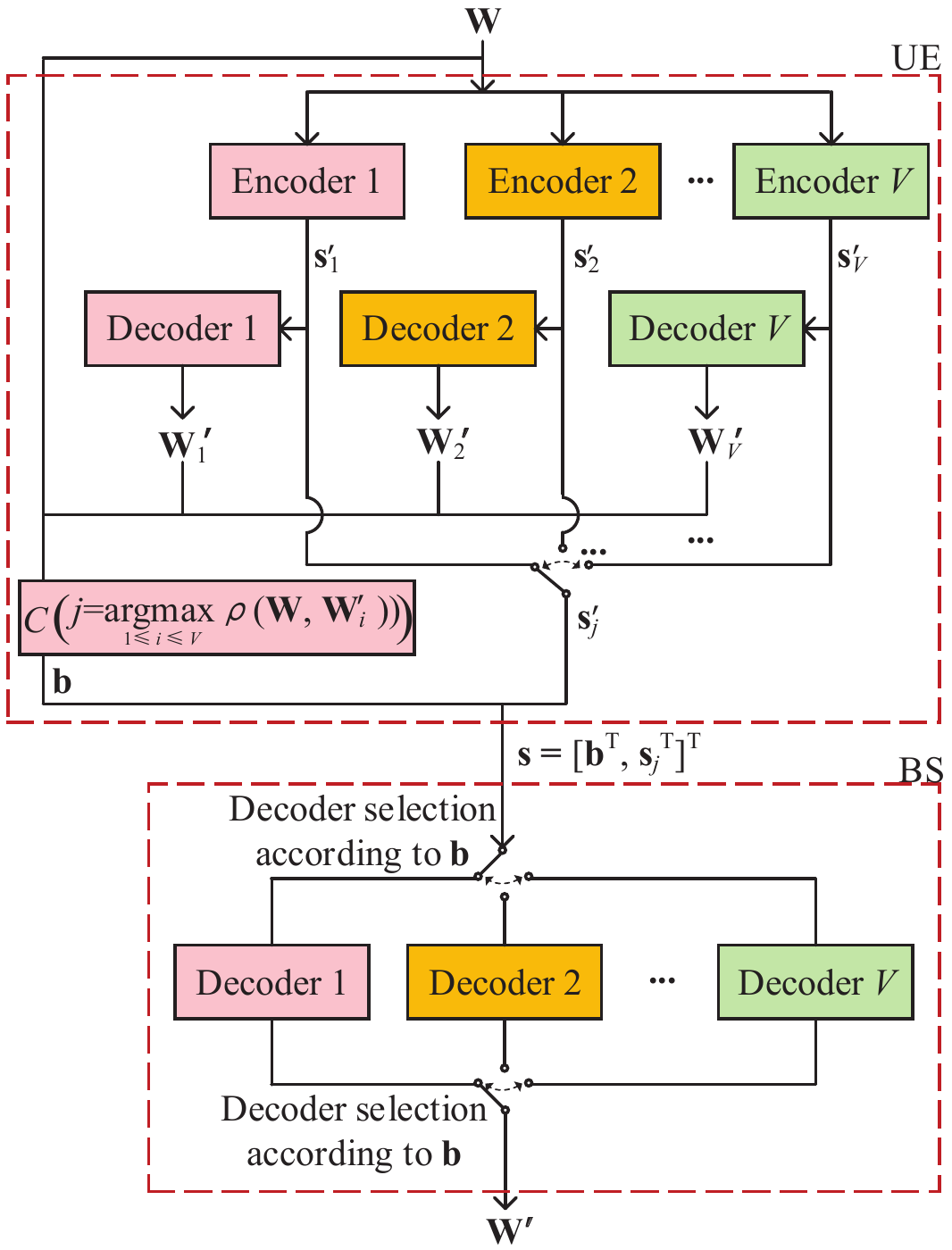}
\caption{Illustration of the method for multi-model ensembling.}
\label{modelEnsembling}
\end{figure}

The model ensemble is another optional enhancing method, which can integrate the ability of multiple models and improve the accuracy performance of the ensemble model during training or testing period. For the method of model ensemble during testing period as an example which is shown in Fig. \ref{modelEnsembling}. $V$ encoder-decoder pairs are constructed with different conditions such as model structure, random seed and batch size for training, etc. The $V$ encoder-decoder pairs are jointly implemented at the UE, wherein the corresponding decoders are also adapted at BS. At the UE side, the CSI $\mathbf{W}$ is firstly processed by $V$ encoder-decoder pairs obtaining encoded bitstreams $\{\mathbf{s}'_1, \dots, \mathbf{s}'_V,\}$ and the decoded CSI $\{\mathbf{W}'_1, \dots, \mathbf{W}'_V \}$, respectively. The decoded CSI can be utilized to select the encoder-decoder pair with the best SGCS performance, whose index is encoded by
\begin{equation}
\begin{split}
\mathbf{b} = C(j = \mathop{\arg\max}\limits_{1 \leq i \leq V} \rho(\mathbf{W}, \mathbf{W}_i'))
\end{split}
\text{,}
\end{equation}
where $j$ denotes the index of best encoder-decoder pair, $C(\cdot)$ denotes index encoding function and $b \in \{0,1\}^{\lceil \log_2^{V} \rceil}$ represents the bitstream indicating the index of the best pair, respectively. The joint bitstream for index and CSI, $\mathbf{s} = [\mathbf{b}^{\rm{T}}, \mathbf{s}_j^{\rm{T}}]^{\rm{T}}$ is fed back to the BS side, where the BS selects the decoder according to $\mathbf{b}$ and decodes the CSI by the selected decoder, obtaining the final recovered CSI $\mathbf{W}'$. The above model ensemble method can combine the advantages of multiple models. However, deploying multiple encoder-decoder pairs on the UE brings storage overhead at the UE side and the feedback overhead for encoding index. Moreover, considering a fixed number of total feedback bits $M = \lceil \log_2^{V} \rceil + M_s$, where $\lceil \log_2^{V} \rceil$ and $M_s$ respectively represent the length of bits overhead for encoding index and CSI tensor, the increasing number of the total models $V$ improves the diversity of ability of the ensemble model, but decreases the length of bits for encoding CSI $M_s$ resulting in lower performance of each single model. Thus the number of models $V$ is the hyperparameter that needs to be carefully tuned.

\section{Experiments}
\label{Experiments}
\subsection{SGCS Performance Comparison}
In this section, the experimental results are provided to verify the superiority of the EVCsiNet-T compared with traditional TypeI (with mode 1) and eTypeII (with the number of orthogonal beams $L=2$ and $4$) codebook based methods \cite{2}, where the system parameters are listed in Table \ref{tabChannelSeting}. The i) channel data in 2nd WAIC, and ii) link-level channel data of clustered delay line (CDL) A and C with delay spread of 30 and 300 ns respectively, are considered.
%Note that the enhancing schemes are not provided in the experiments due to the fact that there are diverse factors affect the performance such as different enhancing schemes, dataset, model design, etc. 
Moreover, the embedding dimension, the number of basic blocks, quantization bits, heads in multi-head attention layer and scaling factor of hidden layer of EVCsiNet-T are set to $N_{\rm e} = 512$, $N_{\rm b = 10}$, $B = 2$, $N_{\rm head} = 16$ and $k_{\rm h} = 2$, respectively. The cosine similarity loss function and the default adaptive momentum (Adam) optimizer with the learning rate of 0.001 are adopted to train the EVCsiNet-T for 300 epochs. 

%\begin{table*}[tb]\small
%\centering
%\caption{SGCS performance comparison on different channel models}
%\label{SGCS1}
%\renewcommand\arraystretch{1.2}
%\setlength{\tabcolsep}{8mm}{
%\begin{tabular}{c|c|c|c|c}
%\hline
% \multirow{2}{*}{Solution}      & \multirow{2}{*}{Feedback bits $M$}          & \multicolumn{3}{c}{SGCS performance $\rho$} \\
% \cline{3-5} & & CDL-A30 & CDL-C300  & UMa \\ \hline
% TypeI     &   32                  & 0.796    & 0.640    & 0.548 \\ \hline
% EVCsiNet-T       & 32                    &  $\bold{0.962}$   & $\bold{0.903}$       & $\bold{0.791}$  \\ \hline
% eTypeII ($L=2$)  &   49                  &  0.906   & 0.785   & 0.702   \\ \hline
% EVCsiNet-T     & 48                    &  $\bold{0.975}$   & $\bold{0.927}$      &  $\bold{0.843}$ \\ \hline
% eTypeII ($L=4$)  & 128                 &  0.951   & 0.868   & 0.853   \\ \hline
% EVCsiNet-T      &120                    &  $\bold{0.987}$   & $\bold{0.967}$    & $\bold{0.914}$ \\ \hline
%\end{tabular}}
%\end{table*}

\begin{figure}[tb]
\centering
\includegraphics[scale=0.68]{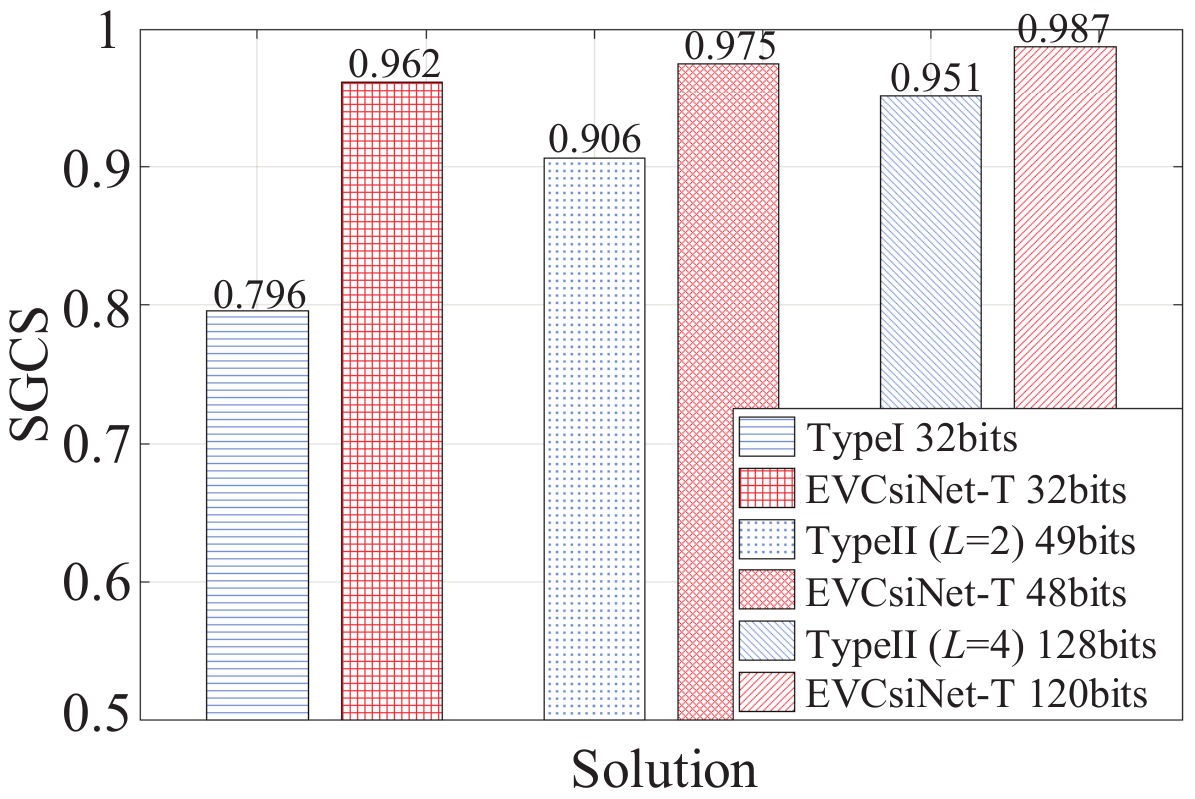}
\caption{SGCS comparison on CDL-A30 for different solutions.}
\label{CDLA}
\end{figure}

\begin{figure}[tb]
\centering
\includegraphics[scale=0.68]{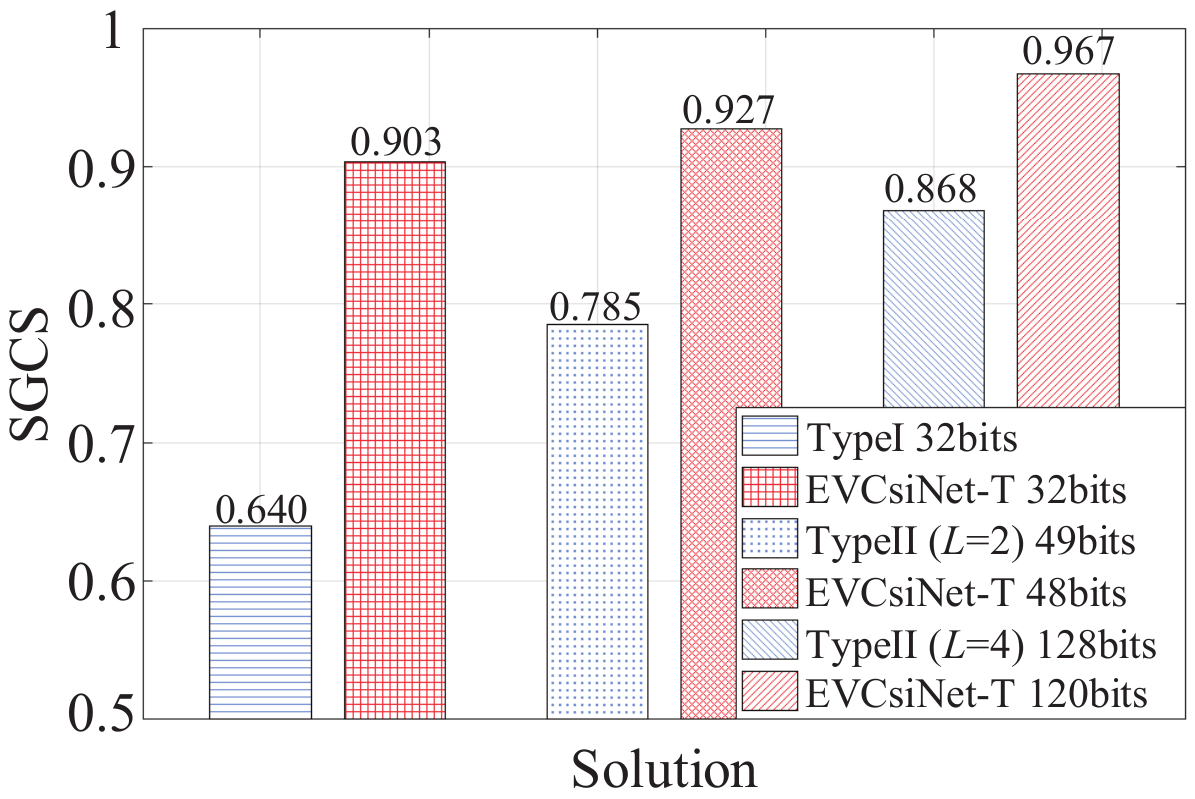}
\caption{SGCS comparison on CDL-C300 for different solutions.}
\label{CDLC}
\end{figure}

\begin{figure}[tb]
\centering
\includegraphics[scale=0.68]{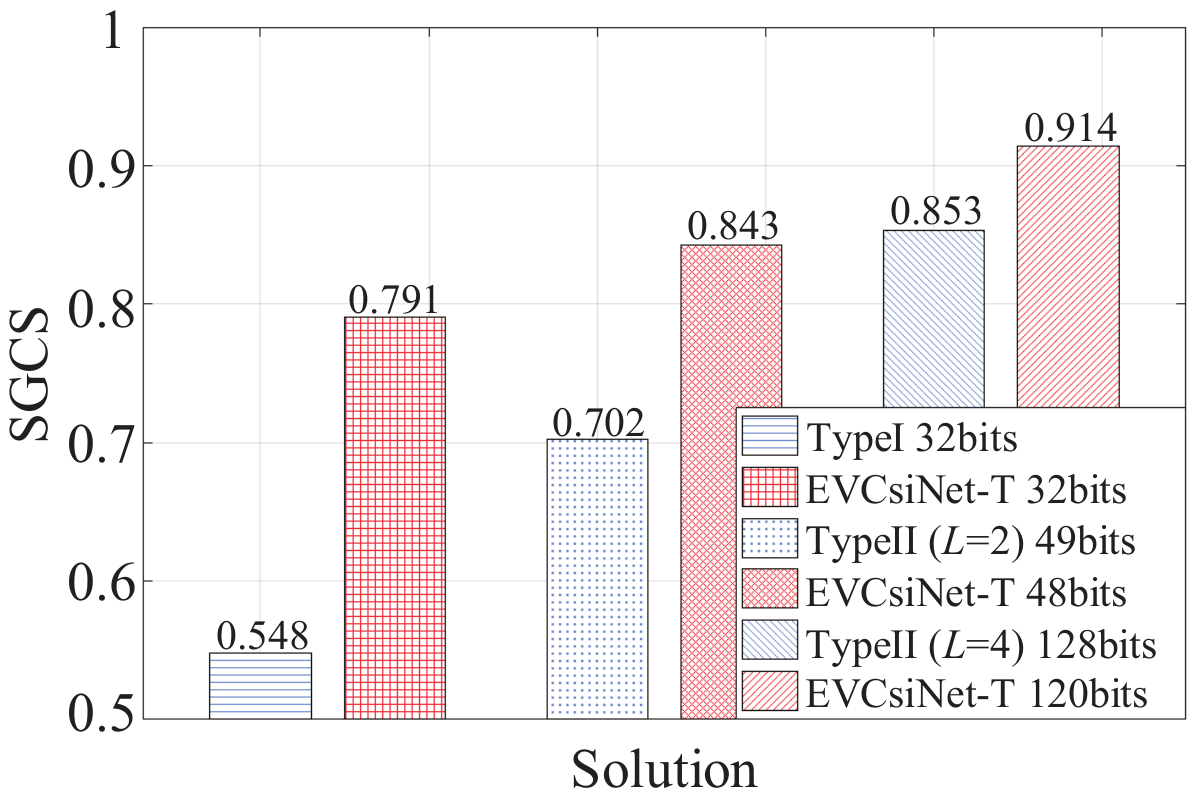}
\caption{SGCS comparison on the dataset in 2nd WAIC for different solutions.}
\label{UMA}
\end{figure}

Fig. \ref{CDLA}, \ref{CDLC} and \ref{UMA} show the SGCS performance comparison between EVCsiNet-T and traditional codebook based TypeI and eTypeII methods over different channels including CDL-A with delay spread of 30 ns (CDL-A30), CDL-C with delay spread of 300 ns (CDL-C300), and the channel in 2nd WAIC. Note that to reveal the superiority of the backbone itself the enhancing methods described in Section \ref{Enhancing Schemes} are not implemented in basic EVCsiNet-T in Fig. \ref{CDLA}, \ref{CDLC} and \ref{UMA}. It can be noticed that under low feedback overhead condition the EVCsiNet-T with $M=32$ and $M=48$ outperform TypeI with $M=32$ and eTypeII ($L=2$) with $M=49$. As for the configuration of high feedback overhead, the EVCsiNet-T with $M=120$ also performs higher SGCS than eTypeII ($L=4$) with $M=128$. Moreover, taking CDL-C300 as an example, the EVCsiNet-T with $M=32$ achieves $\rho = 0.903$ which even outperforms the $\rho = 0.868$ of eTypeII ($L=4$) with higher feedback bits of $M=128$. This indicates that the EVCsiNet-T can significantly reduce the feedback overhead without the cost of performance loss in comparison with traditional codebook based counterparts. Meanwhile, three different types of channels lead to different correlation features between subbands in the frequency domain. This property can be well exploited by EVCsiNet-T for CSI compression and recovery since the different subbands are treated as a sequence of inputs so that the cross-correlated features can be extracted effectively by attention mechanism, which further brings performance gain in comparison with the traditional codebook solution with relative bias in the selection of broadband beam group and subband beams.

\begin{table*}[tb]
\Large
\caption{Transformer based solutions with different enhancing schemes.}
\begin{threeparttable}
\resizebox{\textwidth}{!}{
\renewcommand{\arraystretch}{1}
\begin{tabular}{|m{0.8cm}<{\centering}|m{1.6cm}<{\centering}|m{1.6cm}<{\centering}|m{1.9cm}<{\centering}|m{1.9cm}<{\centering}|m{1.9cm}<{\centering}|m{1.9cm}<{\centering}|m{1.9cm}<{\centering}|m{1.9cm}<{\centering}|m{1.9cm}<{\centering}|m{1.9cm}<{\centering}|m{1.9cm}<{\centering}|}
\hline
\multirow{3}{*}{\begin{tabular}[m]{@{}c@{}}\\ \\Solu-\\ tion\\ ID\end{tabular}} &\multirow{3}{*}{\begin{tabular}[c]{@{}c@{}}\\ \\SGCS \\(48bits)\end{tabular}}&\multirow{3}{*}{\begin{tabular}[c]{@{}c@{}}\\ \\SGCS \\(128bits)\end{tabular}} &\multicolumn{9}{c|}{Scheme}\\ \cline{4-12}
&  & &\multicolumn{4}{c|}{Data Augmentation}                                                                                                                                                                                                                              &\multicolumn{2}{c|}{Loss Function Design}                                                                                                                                                                                                                               &\multicolumn{2}{c|}{Training Strategy}                                                                                                                                                                                                                                &\multirow{2}{*}{\begin{tabular}[c]{@{}c@{}}\\ \\ Model \\Ensemble\end{tabular}}  \\ \cline{4-11}
&  & &Noise Injection  & Flipping &Shifting  &Rotation  & Scoring Loss &  Quan. Error Compensation  & Learning Rate Warm-up and Decay &Staged Training & \\ \hline
%\cline{4-12}
%&&& \multicolumn{4}{c|}{Data Preprocessing}                                                                                                                                                                                                                             & \multicolumn{2}{c|}{Backbone}   & \multicolumn{2}{c|}{Quantization Enhancement}                                                                                                                                                                                                                                                                                                                                                &\\ \cline{4-12}
1 & 0.877 & 0.938&  &  &  &  & $\surd$ &    &$\surd$  & &  $\surd$ \\ \hline
2 & 0.865 & 0.935&  &  &  &  & $\surd$ &  $\surd$  & $\surd$ & &  $\surd$   \\ \hline
3 & 0.850 & 0.930&  &  &  &  & $\surd$ &  $\surd$  &  &$\surd$ &   \\ \hline
4 & 0.849 & 0.922&  & $\surd$ & $\surd$ & $\surd$ & $\surd$ &    & $\surd$ & &   \\ \hline
5 & 0.843 & 0.922& $\surd$ &  & $\surd$ & $\surd$ & $\surd$ &    &$\surd$  &$\surd$ &   \\ \hline
6 & 0.842 & 0.921&  &  &  &  &  &   $\surd$ &  &$\surd$  &  \\ \hline

\end{tabular}
}
%\begin{tablenotes}
%    \normalsize \item[1] Note that the number of feedback bits satisfies that $\rm{NMSE} = \rm{E}\left\{||\mathbf{H}'-\mathbf{H}||_2^2/||\mathbf{H}||_2^2\right\}\leq0.1$.
%\end{tablenotes}
\end{threeparttable}
\label{Capability composition of participating models table}
\end{table*}

Beyond the superior performance of EVCsiNet-T compared to the traditional codebook based methods, there are also some enhancing schemes which can empower EVCsiNet-T and further improve the performance of this backbone. For solving DL-based CSI feedback, in 2nd WAIC there are six over top ten teams use this backbone with various of enhancing schemes introduced in Section \ref{Enhancing Schemes}. Note that since the effectiveness of enhancing scheme is different in the case of different datasets, backbone hyperparameters and joint implementation of enhancing schemes, the simple evaluation of single enhancing scheme is relatively unilateral. Therefore, in Table \ref{Capability composition of participating models table} we also provide the evaluation of good-performing solutions using various combinations of enhancement schemes on the dataset in 2nd WAIC setting the number of feedback bits $M=48$ and $128$, and check mark indicates the utilization of corresponding enhancing scheme.

\begin{figure}[tb]
\centering
\includegraphics[scale=0.47]{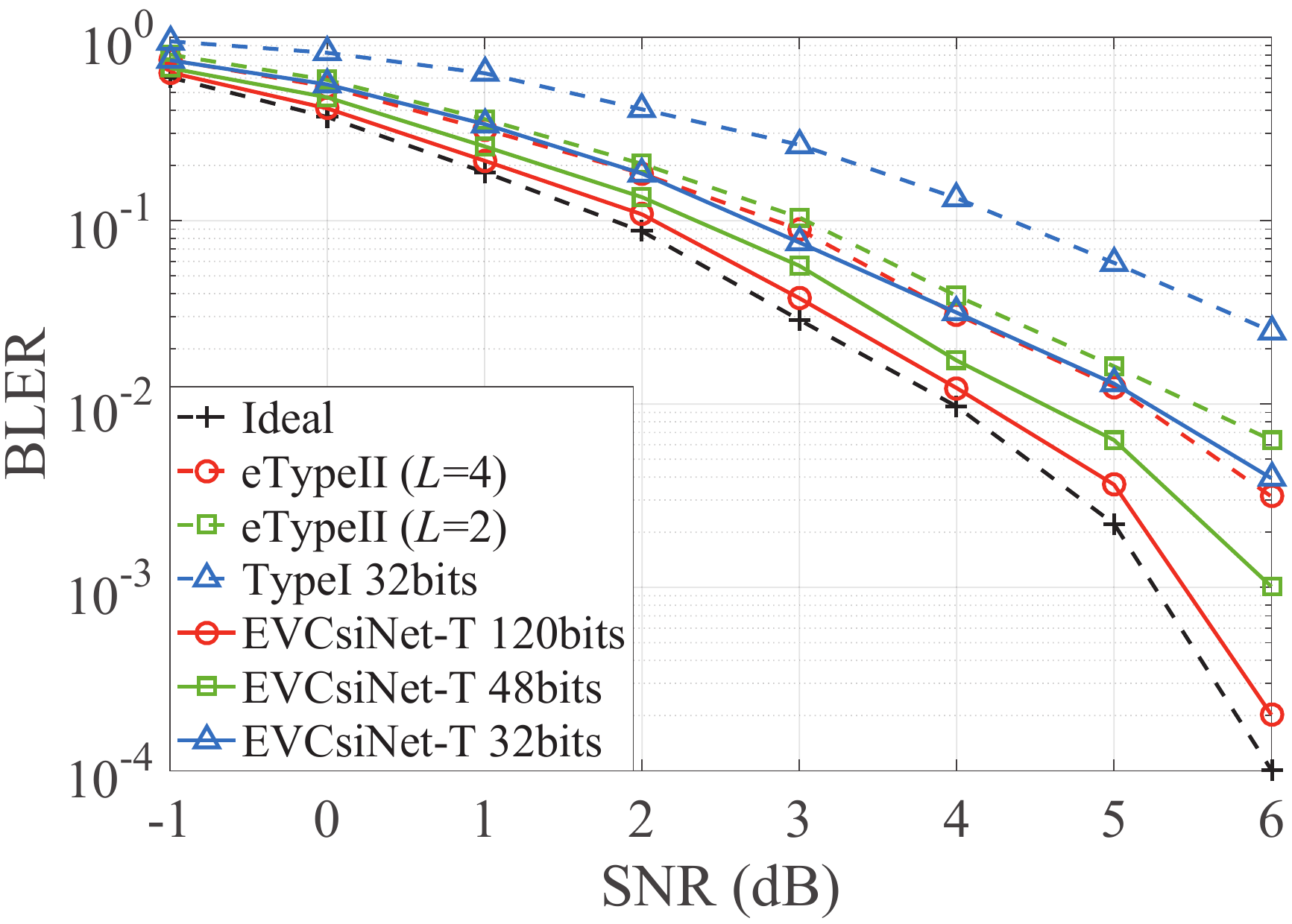}
\caption{Link-level BLER performance comparison on CDL-C300 for
different solutions..}
\label{BLER}
\end{figure}
\subsection{BLER Performance Comparison}
As shown in Fig. \ref{BLER} to further reveal the effectiveness of the proposed method, the link-level block error rate (BLER) performace comparison on CDL-C300 channels for different solutions is provided with the signal-to-noise (SNR) from -1 to 6 dB over $10^{4}$ realizations. Moreover, without considering adaptive modulation and coding, the modulation and coding scheme (MCS) is configured as 19 \cite{2}. The ideal channel estimation is assumed where the BS knows the perfect CSI with SGCS = 1. We can notice that the proposed EVCsiNet-T with $M = 120$ can achieve BLER  \textless $10^{-3}$ with SNR = 6 dB and significantly outperforms eTypeII of $L = 4$ with overhead of 128 bits. Moreover, the EVCsiNet-T method can greatly reduce the feedback overhead with similar BLER performance, since the EVCsiNet-T with $M = 48$ and $32$ slightly outperform eTypeII $L=4$ and $2$ with feedback bits of 128 and 49, respectively.

\subsection{Complexity Analysis}

The complexity analysis of EVCsiNet-T with the number of feedback bits $M = 32, 48$ and $120$ is provided in Table \ref{tabFLOPs} from the perspective of floating point operations (FLOPs) and trainable parameters. 
\begin{table}[!ht]
\small
\centering
\caption{FLOPs and number of trainable parameters.}
\label{tab4}
\setlength{\tabcolsep}{2.4mm}{
%\vspace{-3mm}
{\begin{tabular}{|c|c|c|c|c|}
\hline
\multirow{2}{*}{$M$}      & \multicolumn{2}{c|}{FLOPs ($\times 10^{7}$) } &  \multicolumn{2}{c|}{Trainable Par. ($\times 10^{7}$) }\\ 
 \cline{2-5} & Encoder & Decoder &  Encoder & Decoder   \\ \hline
$32$   & $4.2099$             & 4.2099 & $2.1107$ & 2.1108 \\ \hline
$48$  & 4.2111             & 4.2111 & 2.1113 & 2.1114 \\ \hline
$120$   & 4.2166           & 4.2166 & 2.1141 & 2.1142 \\ \hline
\end{tabular}}
\label{tabFLOPs}}
\end{table}

Obviously, the EVCsiNet-T with $M=32$ requires $8.4198 \times 10^7$ FLOPs for inference. Since the feedback delay of the existing communication system is required to be less than 1 ms, the inference time using a common device of NVIDIA Tesla V100 SXM2 with double-precision performance of  $7.8 \times 10^{12}$ floating point operations per second (FLOPS) can theoretically be 10.8 $\mu$s which meets the requirement well. It can also be noticed that the FLOPs and trainable parameters are quite close for different number of feedback bits, which indicates that the the inference time can not be affected too much by different number of feedback bits.

\section{Conclusion}
\label{CONCLUSION}

In this paper, we first give a description of the framework of CSI feedback and its corresponding channel model involved in the 2nd WAIC. Then a Transformer backbone for CSI feedback referred to EVCsiNet-T is proposed. Moreover, a series of enhancement schemes in 2nd WAIC including i) data augmentation, ii) loss function design, iii) training strategy, and iv) model ensemble are introduced. The experimental results involving the comparison between EVCsiNet-T and traditional codebook methods over different channels are further provided, which show the advanced performance and a promising prospect of Transformer on CSI feedback problem. Since the significance and indispensability of interpretable ML in the AI-enhanced wireless communication, we will study on this field with more theoretical interpretation and analysis in future research. Finally we sincerely hope that this article and 2nd WAIC can broaden the thinking and insights for interested researchers.

\section*{Acknowledgement}
\label{ACKNOWLEDGEMENT}
We sincerely thank China Academy of Information and Communications Technology, Guangdong OPPO Mobile Telecommunications Corp., Ltd, National Mobile
Communications Research Laboratory of Southeast University and vivo Mobile Communication Co., Ltd for their great help and support on the research and the holding of the 2nd WAIC. We also would like to express our gratitude to all contestants for their participation and sharing.

\bibliographystyle{gbt7714-numerical}
\bibliography{main}

\begin{thebibliography}{30}
\providecommand{\natexlab}[1]{#1}
\providecommand{\url}[1]{#1}
\expandafter\ifx\csname urlstyle\endcsname\relax\else
  \urlstyle{same}\fi
\expandafter\ifx\csname href\endcsname\relax
  \DeclareUrlCommand\doi{\urlstyle{rm}}
  \def\eprint#1#2{#2}
\else
  \def\doi#1{\href{https://doi.org/#1}{\nolinkurl{#1}}}
  \let\eprint\href
\fi

\bibitem[3GPP(2020{\natexlab{a}})]{1}
3GPP.
\newblock {3GPP TS} 38.212 v16.1.0, 3rd generation partnership project;
  technical specification group radio access network; {NR}; multiplexing and
  channel coding (release 16)[M].
\newblock [S.l.: s.n.], 2020{\natexlab{a}}.

\bibitem[3GPP(2020{\natexlab{b}})]{2}
3GPP.
\newblock {3GPP TS} 38.214 v16.1.0 3rd generation partnership project;
  technical specification group radio access network; {NR}; physical layer
  procedures for data (release 16)[M].
\newblock [S.l.: s.n.], 2020{\natexlab{b}}.

\bibitem[3GPP(2020{\natexlab{c}})]{3}
3GPP.
\newblock {3GPP TS} 38.331 v16.0.0 3rd generation partnership project;
  technical specification group radio access network; {NR}; radio resource
  control ({RRC}) protocol specification (release 16)[M].
\newblock [S.l.: s.n.], 2020{\natexlab{c}}.

\bibitem[Xiao et~al.(2021)Xiao, Wang, Tian, Liu, Liu, Jin, Shen, Zhang, and
  Yang]{Xiao2021AIEW}
XIAO~H, WANG~Z, TIAN~W, et~al.
\newblock {AI enlightens wireless communication: Analyses, solutions and
  opportunities on CSI feedback}[J].
\newblock China Communications, 2021, 18: 104-116.

\bibitem[Xiao et~al.(2022)Xiao, Tian, Liu, and Shen]{xiao2022channelgan}
XIAO~H, TIAN~W, LIU~W, et~al.
\newblock {ChannelGAN: Deep learning based channel modeling and generating}[J].
\newblock IEEE Wireless Communications Letters, 2022.

\bibitem[Liu et~al.(2021)Liu, Tian, Xiao, Jin, Liu, and Shen]{liu2021evcsinet}
LIU~W, TIAN~W, XIAO~H, et~al.
\newblock {EVCsiNet: Eigenvector-based CSI feedback under 3GPP link-level
  channels}[J].
\newblock IEEE Wireless Communications Letters, 2021, 10\allowbreak (12):
  2688-2692.

\bibitem[Qualcomm(2021)]{213599}
QUALCOMM.
\newblock {RP-213599, New SI: Study on Artificial Intelligence (AI)/Machine
  Learning (ML) for NR Air Interface}[C]//\allowbreak
{3GPP TSG RAN{\#}94e, electronic meeting}.
\newblock [S.l.]: 3GPP, 2021.

\bibitem[OPPO(2021)]{212927}
OPPO.
\newblock {RP-212927, Discussion on R18 study on AIML-based 5G
  enhancements}[C]//\allowbreak
{3GPP TSG RAN{\#}94e, electronic meeting}.
\newblock [S.l.]: 3GPP, 2021.

\bibitem[CAICT(2021{\natexlab{b}})]{210235}
CAICT.
\newblock {RWS-210235, Views on AI based physical layer
  enhancements}[C]//\allowbreak
{3GPP TSG RAN Rel-18 Workshop, electronic meeting}.
\newblock [S.l.]: 3GPP, 2021{\natexlab{b}}.

\bibitem[CAICT et~al.(2021{\natexlab{a}})CAICT and OPPO]{210236}
CAICT, OPPO.
\newblock {RWS-210236, Introduction of the 1st wireless communication AI
  competition (WAIC)}[C]//\allowbreak
{3GPP TSG RAN Rel-18 Workshop, Electronic Meeting}.
\newblock [S.l.]: 3GPP, 2021{\natexlab{a}}.

\bibitem[Wang et~al.(2017)Wang, Wen, Wang, Gao, Jiang, and Jin]{wang2017deep}
WANG~T, WEN~C~K, WANG~H, et~al.
\newblock Deep learning for wireless physical layer: opportunities and
  challenges[J].
\newblock China Communications, 2017, 14\allowbreak (11): 92-111.

\bibitem[Wen et~al.(2018)Wen, Shih, and Jin]{wen2018deep}
WEN~C~K, SHIH~W~T, JIN~S.
\newblock Deep learning for massive {MIMO} {CSI} feedback[J].
\newblock IEEE Wireless Communications Letters, 2018, 7\allowbreak (5):
  748-751.

\bibitem[Sun et~al.(2020)Sun, Xu, Fan, Li, and Karagiannidis]{sun2020ancinet}
SUN~Y, XU~W, FAN~L, et~al.
\newblock Ancinet: An efficient deep learning approach for feedback compression
  of estimated {CSI} in massive {MIMO} systems[J].
\newblock IEEE Wireless Communications Letters, 2020, 9\allowbreak (12):
  2192-2196.

\bibitem[Lu et~al.(2020)Lu, Wang, and Song]{lu2020multi}
LU~Z, WANG~J, SONG~J.
\newblock Multi-resolution {CSI} feedback with deep learning in massive {MIMO}
  system[C]//\allowbreak
ICC 2020-2020 IEEE International Conference on Communications (ICC).
\newblock [S.l.]: IEEE, 2020: 1-6.

\bibitem[Chen et~al.(2020)Chen, Guo, Wen, Jin, Li, Wang, and Hou]{chen2020deep}
CHEN~T, GUO~J, WEN~C~K, et~al.
\newblock Deep learning for joint channel estimation and feedback in massive
  {MIMO} systems[J].
\newblock arXiv preprint arXiv:2011.07242, 2020.

\bibitem[Mashhadi et~al.(2020)Mashhadi, Yang, and
  G{\"u}nd{\"u}z]{mashhadi2020distributed}
MASHHADI~M~B, YANG~Q, G{\"U}ND{\"U}Z~D.
\newblock Distributed deep convolutional compression for massive {MIMO} {CSI}
  feedback[J].
\newblock IEEE Transactions on Wireless Communications, 2020.

\bibitem[Cao et~al.(2021)Cao, Shih, Guo, Wen, and Jin]{cao2021lightweight}
CAO~Z, SHIH~W~T, GUO~J, et~al.
\newblock Lightweight convolutional neural networks for {CSI} feedback in
  massive {MIMO}[J].
\newblock IEEE Communications Letters, 2021.

\bibitem[Guo et~al.(2020{\natexlab{a}})Guo, Wen, and Jin]{guo2020deep}
GUO~J, WEN~C~K, JIN~S.
\newblock Deep learning-based {CSI} feedback for beamforming in single-and
  multi-cell massive {MIMO} systems[J].
\newblock IEEE Journal on Selected Areas in Communications, 2020{\natexlab{a}}.

\bibitem[Guo et~al.(2021)Guo, Wen, and Jin]{guo2021canet}
GUO~J, WEN~C~K, JIN~S.
\newblock Canet: Uplink-aided downlink channel acquisition in {FDD} massive
  {MIMO} using deep learning[J].
\newblock arXiv preprint arXiv:2101.04377, 2021.

\bibitem[Lu et~al.(2018)Lu, Xu, Shen, Zhu, and Wang]{lu2018mimo}
LU~C, XU~W, SHEN~H, et~al.
\newblock {MIMO} channel information feedback using deep recurrent network[J].
\newblock IEEE Communications Letters, 2018, 23\allowbreak (1): 188-191.

\bibitem[Wang et~al.(2018)Wang, Wen, Jin, and Li]{wang2018deep}
WANG~T, WEN~C~K, JIN~S, et~al.
\newblock Deep learning-based {CSI} feedback approach for time-varying massive
  {MIMO} channels[J].
\newblock IEEE Wireless Communications Letters, 2018, 8\allowbreak (2):
  416-419.

\bibitem[Guo et~al.(2020{\natexlab{b}})Guo, Wen, Jin, and
  Li]{guo2020convolutional}
GUO~J, WEN~C~K, JIN~S, et~al.
\newblock Convolutional neural network-based multiple-rate compressive sensing
  for massive {MIMO} {CSI} feedback: design, simulation, and analysis[J].
\newblock IEEE Transactions on Wireless Communications, 2020{\natexlab{b}},
  19\allowbreak (4): 2827-2840.

\bibitem[Li et~al.(2020)Li and Wu]{li2020spatio}
LI~X, WU~H.
\newblock Spatio-temporal representation with deep neural recurrent network in
  {MIMO} {CSI} feedback[J].
\newblock IEEE Wireless Communications Letters, 2020, 9\allowbreak (5):
  653-657.

\bibitem[Chen et~al.(2019)Chen, Guo, Jin, Wen, and Li]{chen2019novel}
CHEN~T, GUO~J, JIN~S, et~al.
\newblock A novel quantization method for deep learning-based massive {MIMO}
  {CSI} feedback[C]//\allowbreak
2019 IEEE Global Conference on Signal and Information Processing (GlobalSIP).
\newblock [S.l.]: IEEE, 2019: 1-5.

\bibitem[Lu et~al.(2019)Lu, Xu, Jin, and Wang]{lu2019bit}
LU~C, XU~W, JIN~S, et~al.
\newblock Bit-level optimized neural network for multi-antenna channel
  quantization[J].
\newblock IEEE Wireless Communications Letters, 2019, 9\allowbreak (1): 87-90.

\bibitem[Vaswani et~al.(2017)Vaswani, Shazeer, Parmar, Uszkoreit, Jones, Gomez,
  Kaiser, and Polosukhin]{Vaswani2017AttentionIA}
VASWANI~A, SHAZEER~N~M, PARMAR~N, et~al.
\newblock Attention is all you need[J].
\newblock ArXiv, 2017, abs/1706.03762.

\bibitem[Dosovitskiy et~al.(2021)Dosovitskiy, Beyer, Kolesnikov, Weissenborn,
  Zhai, Unterthiner, Dehghani, Minderer, Heigold, Gelly, Uszkoreit, and
  Houlsby]{Dosovitskiy2021AnII}
DOSOVITSKIY~A, BEYER~L, KOLESNIKOV~A, et~al.
\newblock An image is worth 16x16 words: Transformers for image recognition at
  scale[J].
\newblock ArXiv, 2021, abs/2010.11929.

\bibitem[3GPP(2020{\natexlab{d}})]{4}
3GPP.
\newblock {3GPP TR} 38.901 v16.1.0 3rd generation partnership project;
  technical specification group radio access network; study on channel model
  for frequencies from 0.5 to 100 {GH}z (release 16)[M].
\newblock [S.l.: s.n.], 2020{\natexlab{d}}.

\bibitem[You et~al.(2019)You, Long, Jordan, and Jordan]{You2019HowDL}
YOU~K, LONG~M, JORDAN~M~I, et~al.
\newblock How does learning rate decay help modern neural networks[J].
\newblock arXiv: Learning, 2019.

\bibitem[Liang et~al.(2021)Liang, Zharkov, Amjadi, Joze, and
  Pradeep]{Liang2021GuidanceNW}
LIANG~L, ZHARKOV~I, AMJADI~F, et~al.
\newblock Guidance network with staged learning for image enhancement[J].
\newblock 2021 IEEE/CVF Conference on Computer Vision and Pattern Recognition
  Workshops (CVPRW), 2021: 836-845.

\end{thebibliography}

\newpage
\biographies

\end{document}